\documentclass{article}

\usepackage{arxiv}

\usepackage[utf8]{inputenc} 
\usepackage[T1]{fontenc}    
\usepackage{hyperref}       
\usepackage{url}            
\usepackage{booktabs}       
\usepackage{amsfonts}       
\usepackage{nicefrac}       
\usepackage{microtype}      
\usepackage{lipsum}
\usepackage{color}

\usepackage{multicol}
\usepackage{multirow}
\usepackage{threeparttable}
\usepackage{enumitem}
\usepackage{tikz}
\usepackage{amsmath}
\usepackage{amssymb}

\newcommand{\tts}[1]{\small\ttfamily}
\newcommand{\greendot}[2][orange,fill=orange]{\tikz[baseline=-0.5ex]\draw[#1,radius=#2] (0,0) circle ;}%
\title{Predicting the Reproducibility of Social and Behavioral Science Papers Using Supervised Learning Models}

\author{
  Jian~Wu\\
  Computer Science\\
  Old Dominion University \\
  Norfolk, VA, USA\\
  \texttt{jwu@cs.odu.edu} \\
  \And
  Rajal Nivargi \\
  Information Sciences and Technology\\
  Pennsylvania State University\\
  University Park, PA, USA\\
  \texttt{rfn5089@psu.edu}\\
  \And
  Sree Sai Teja Lanka\\
  Information Sciences and Technology\\
  Pennsylvania State University\\
  University Park, PA, USA\\
  \texttt{szl577@psu.edu}\\
  \AND
  Arjun Manoj Menon\\
  Information Sciences and Technology\\
  Pennsylvania State University\\
  University Park, PA, USA\\
  \texttt{arjunmenon@psu.edu}\\
  \And
  Sai Ajay Modukuri\\
  Information Sciences and Technology\\
  Pennsylvania State University\\
  University Park, PA, USA\\
  \texttt{svm6277@psu.edu}\\
  \And
  Nishanth Nakshatri\\
  Information Sciences and Technology\\
  Pennsylvania State University\\
  University Park, PA, USA\\
  \texttt{nzn5185@psu.edu}\\
  \And
  Xin Wei\\
  Computer Science\\
  Old Dominion University\\
  Norfolk, VA, USA\\
  \texttt{nzn5185@psu.edu}
  \And
  Zhuoer Wang\\
  Computer Science and Engineering\\
  Texas A\&M University\\
  College Station, TX, USA\\
  \texttt{wang@tamu.edu}
  \AND
  James Caverlee\\
  Computer Science and Engineering\\
  Texas A\&M University\\
  College Station, TX, USA\\
  \texttt{caverlee@tamu.edu}
  \And
  Sarah M. Rajtmajer\\
  Information Sciences and Technology\\
  Pennsylvania State University\\
  University Park, PA, USA\\
  \texttt{smr48@psu.edu}
  \And
  C. Lee Giles\\
  Information Sciences and Technology\\
  Pennsylvania State University\\
  University Park, PA, USA\\
  \texttt{giles@ist.psu.edu}
}

\begin{document}
\maketitle

\begin{abstract}
In recent years, significant effort has been invested verifying the reproducibility and robustness of research claims in social and behavioral sciences (SBS), much of which has involved resource-intensive replication projects. In this paper, we investigate prediction of the reproducibility of SBS papers using machine learning methods based on a set of features. We propose a framework that extracts five types of features from scholarly work that can be used to support assessments of reproducibility of published research claims. Bibliometric features, venue features, and author features are collected from public APIs or extracted using open source machine learning libraries with customized parsers. Statistical features, such as p-values, are extracted by recognizing patterns in the body text. Semantic features, such as funding information, are obtained from public APIs or are extracted using natural language processing models. We analyze pairwise correlations between individual features and their importance for predicting a set of human-assessed ground truth labels. In doing so, we identify a subset of 9 top features that play relatively more important roles in predicting the reproducibility of SBS papers in our corpus. Results are verified by comparing performances of 10 supervised predictive classifiers trained on different sets of features.
\end{abstract}


\section{Introduction}
Reproducibility is a defining principle of empirical science. Trust in scientific claims should be based on the robustness of the evidence (e.g., experiments) supporting these claims and subject to verification. 
Concerns about the reproducibility of published research have gained widespread attention over the past decade, with particular focus on the social and behavioral sciences (SBS), e.g.,  \cite{baker2016nature,osc2015estimating,dreber2015using,camerer2016science,klein2018many,camerer2018evaluating,forsell2019predicting}.
By SBS here, we include major disciplines chiefly focused on understanding human behaviors and social systems, such as sociology, political science, economics, and psychology, as well as their respective sub-disciplines, such as computational social science, behavioral economics, social psychology, etc. Research that appears confirmatory when it is in fact exploratory, or makes assertions or predictions that are unlikely given known practical and theoretical limitations, can lead the scientific community and its stakeholders to have inappropriate confidence in reported findings. Manually verifying the reproducibility of scientific claims is non-trivial, usually involving collaboration with original authors and revisiting original experiments, e.g., \cite{camerer2016science}. These concerns have resulted in a call to arms for greater transparency and openness throughout the research process \cite{nosek2016transparency, nosek2018preregistration, munafo2017manifesto}, increased attention to statistical rigor \cite{benjamin2018redefine, wasserstein2016asa}, and even the development of automated approaches to assess confidence in existing claims.\footnote{E.g., DARPA's Systematizing Confidence in Open Research and Evidence (SCORE) program.} The work we describe here aims to support the development of computational tools to assist human understanding of the reproducibility, replicability and generalizability of published claims in the SBS literatures.  

Specifically, a scientific claim appearing in a research article is a testable, falsifiable assertion typically drawing on existing theories or experiments. In this paper, we focus on claims that represent the authors' final conclusions, drawn based on all studies (theoretical and experimental) in the paper.
Below is an exemplar claim from the abstract of a  paper recently shared in PsyArXiv \cite{stanley2020hoax}. It represents the authors' final conclusions, drawn based on all studies (theoretical and experimental) in the paper.
\begin{quote}
    \tt\small Our results indicate that individuals less willing to engage effortful, deliberative, and reflective cognitive processes were more likely to believe the pandemic was a hoax, and less likely to have recently engaged in social-distancing and hand-washing.
\end{quote}


Traditionally, the task described above can be treated as a classification problem and can be approached using traditional models such as the bag-of-words or sequence tagging models. However, just incorporating semantic information is insufficient for this task due to the limited information existing in the local representation of a claim itself. Rather, indicators of a claim's reproducibility should be extracted beyond the claim to include broader context, including but not limited to the background of the paper's authors, whether and how the claim is supported by preceding work in the literature, and whether and how the claim is accredited by subsequent work. In this work, we focus on the \emph{conclusive claim} of an SBS paper and assume that its reproducibility can be predicted using global features extracted from the paper. The features we extract are not aligned with a specific claim. However, they offer a representation of a paper's overall profile, which can be useful as the big picture before zooming into claim-dependent reproducibility prediction. 
Of note, in this work, we focus on feature extraction rather than the prediction model, so we do not distinguish between reproducibility and replicability \cite{goodman2016science}. We assert that these features can be used for prediction modules in either context, so we use claim reproduciblity to broadly describe the ultimate task.

We propose a software framework that extracts 41 features from SBS papers that can be used for reproducibility assessment. 
Extracted features include five types -- bibliometric, venue-related, author-related, statistical, and semantic. Features are extracted using heuristic, machine learning-based methods, and open source software. Individual feature extractors are tested and evaluated based on manually-annotated ground truth when possible. Here, we focus on describing and evaluating the feature extraction framework and briefly describe the initial utility of these features for reproducibility assessment.

\section{Related Works}

\subsection{Information Extraction Systems}\label{sec:extractionSystems}
GROBID (GeneRation Of BIbliographic Data) is a machine learning library for extracting, parsing and re-structuring raw documents such as PDF into structured XML/TEI encoded documents with a particular focus on technical and scientific publications \cite{lopez2009grobid}. GROBID can be used for extracting metadata from headers, citations, and citation context.  Developed based on a hierarchical conditional random field model, GROBID has shown superior performance over many metadata and citation extraction software packages \cite{lipinski2013,tkaczyk18jcdl}. One feature is to segment the full text of an article to section levels, which helps downstream tasks to quickly locate certain section, such as acknowledgements.

CERMINE (Content ExtRactor and MINEr) is a comprehensive open-source system for extracting structured metadata from scientific articles in a born-digital form \cite{tkaczyk2015cermine}.  CERMINE is based on a modular workflow designed to extract basic structure (e.g., page segmentation), metadata, and bibliographies. Similar to GROBID, CERMINE segments documents into \emph{metadata}, \emph{body}, \emph{references}, and \emph{other}. The implementations were based on heuristic, e.g., metadata extraction, machine learning, e.g., metadata zone classification and reference string extraction, and an external library iText\footnote{{https://itextpdf.com/en}}. In \cite{tkaczyk2015cermine}, the authors compared the performance of various metadata extraction systems, including GROBID, based on three datasets. CERMINE consistently outperforms GROBID in all datasets in terms of title, abstract, year, and references. On certain datasets, CERMINE outperforms GROBID on certain datasets, such as authors and affiliations, keywords. 

PDFMEF (PDF Multi-entity Extraction Framework) is a customizable and scalable framework for extracting multiple types of content from scholarly documents \cite{wu2015pdfmef}. PDFMEF encapsulates various content classifiers and extractors (e.g., PDFBox\footnote{{https://pdfbox.apache.org}}, academic paper classifier \cite{caragea2016aaai}, pdffigures2 \cite{clark2016jcdl_pdffigures2}) for multi-type and scalable information extraction tasks. Users can substitute out-of-box extractors with alternatives. 

{\sc ScienceParse}\footnote{{https://github.com/allenai/science-parse}} is a system to extract structured data from raw academic PDFs. The system was able to extract basic bibliographic header fields from a paper, such as title, authors, and abstract. It can extract and parse citation references and their mentions. It can also segment papers into sections, each with heading and body text. A new version\footnote{\url{https://github.com/allenai/spv2}} works in a completely different way with fewer output fields but higher quality output. 

\subsection{Public APIs for Scholarly Articles}
Elsevier\footnote{https://www.elsevier.com/} is an information analytics company which provides a wealth of useful information from books and journals. Along with providing a web browser user-friendly experience, Elsevier also offers APIs to search and retrieve data from their products in a machine readable manner. In this pipeline, the Scopus\footnote{https://www.scopus.com/} APIs are used to retrieve bibliometric information of the scholarly articles. Scopus is the largest abstract and citation database of peer-reviewed literature. With over 77.8 million records and 25,100 journal titles from more than 5000 international publishers, Scopus provides research metrics in the fields of science, technology, medicine, social science and arts and humanities. The Scopus APIs allows real-time access articles, authors and institutions in their database. 

CrossRef\footnote{https://www.crossref.org/} is a not-for-profit association offering an array of services to ensure that scholarly research metadata is registered, linked, and distributed. It interlinks millions of items from a variety of content types, such as journals, books, conference proceedings, working papers, and technical reports. The metadata collected from the members of Crossref can be accessed using the Crossref Metadata Retrieval API\footnote{\url{https://www.crossref.org/services/metadata-retrieval/}}.

Semantic Scholar\footnote{https://www.semanticscholar.org/} is an AI-backed search engine for academic publications. It is designed to highlight the most important and influential papers, and to identify the connections between them. It provides a RESTful API for linking or articles and extracting information from the records on demand. 

\section{Extraction Framework}
\label{sec:framework}
\begin{figure*}
    \centering
    \includegraphics[width=.9\textwidth]{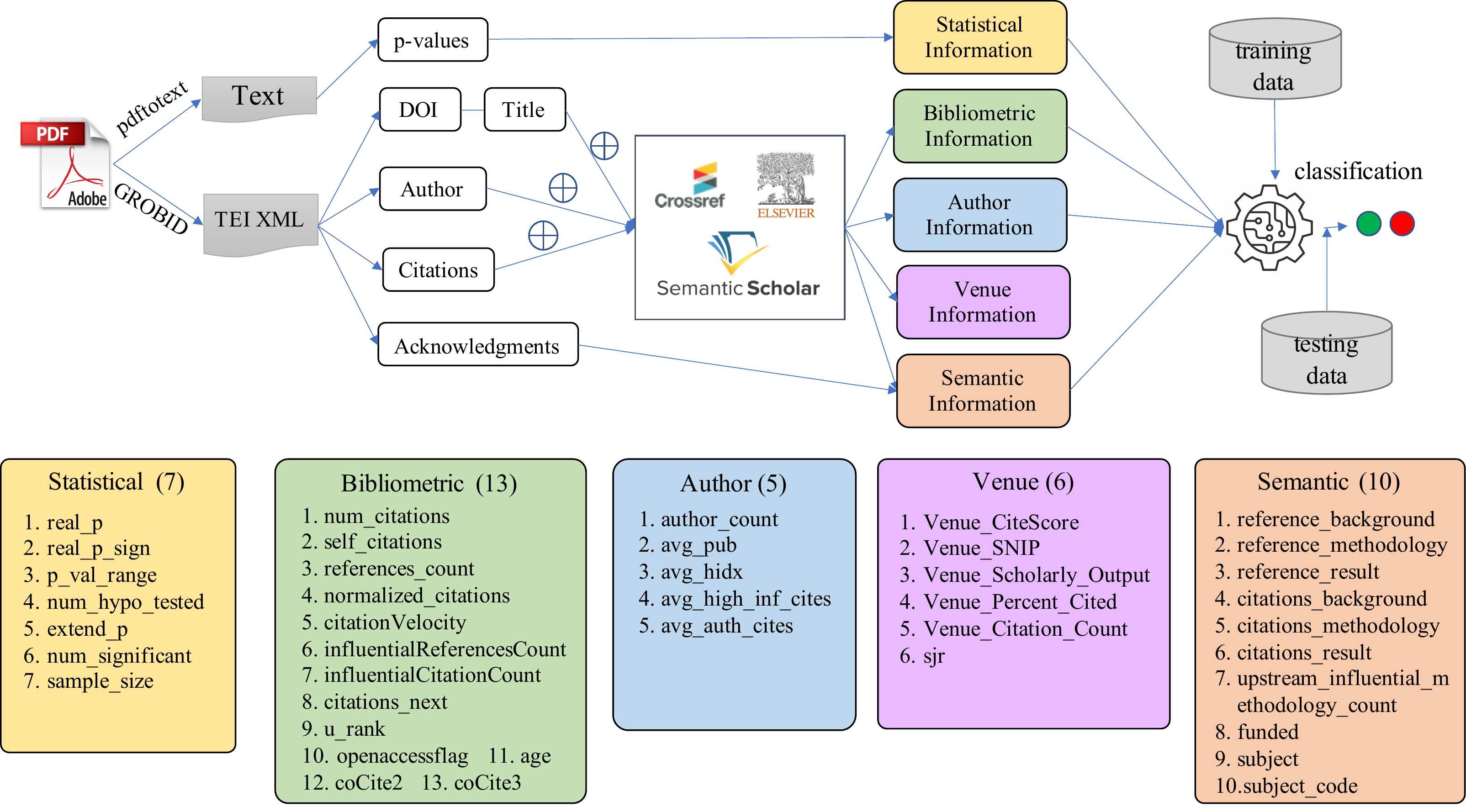}
    \caption{A summary of the feature extraction framework. Different types of features are color-coded.}
    \label{fig:pipelinearch}
\end{figure*}

\subsection{Overview}
The top-level architecture of our extraction framework called FEXRep (Feature EXtraction framework for Replicability prediction) is depicted in Figure~\ref{fig:pipelinearch}. {FEXRep} extracts bibliometric, venue, author,  statistic, and semantic features from SBS papers. The extraction is performed  in two steps. In the first step, we extract raw information directly from content, such as DOI, authors, and citations. In the second step, we derive numerical values to build feature vectors. Our choice of features is based on intuition and promising evidence that certain features may be relevant indicators of reproducibility of findings. We gain deeper insight into their utilities from post-extraction feature analysis.  

\subsection{Document Preprocessing}
The preprocessing step involves parsing unstructured data and generating intermediate metadata that will be used for deriving numerical features. We apply GROBID, which generates a machine-readable XML document under the TEI schema. 
The machine-readable XML output generated is then further parsed to extract entities, such as address, organizations, authors, and their properties. We use {\sc pdftotext} to generate the plain text that is used for extracting p-values. 

\subsection{Bibliometric Features}
This type of features includes quantitative measurements about the target paper's impact in its community.  
For example, well-researched publications are known to feature a thoroughly curated set of references. Poorly curated bibliographies indicate a lack of cohesiveness in arguments or claims, and insufficient effort invested towards validating ideas against existing works. 
Citations made in the context of using or extending methodologies presented in cited work can serve as additional indicators of the robustness of methodology and the reproducibility of claims. 
Thus, assessing the references and citations of an academic publications is potentially an important source that could help discern reproducible research.

A complete listing of bibliometric features extracted is publically available in an online document\footnote{\url{shorturl.at/ghtD2}}. We use the DOI or the title (if DOI is not available) extracted by GROBID as a paper's identifier. Many bibliometric values are obtained by querying digital library APIs, including the Crossref Metadata Retrieval API (hereafter Crossref), Elsevier Scopus API (hereafter Scopus), and Semantic Scholar API\footnote{https://api.semanticscholar.org/} (hereafter referred as  S2). The records from an API response are refined by calculating string similarties between their titles and the title of the queried paper because in some cases, GROBID returns a partial title. The record whose matching score is greater than 90\% is chosen as the final matching result. 

\paragraph{num\_citations\label{sec:citedby}} 
This metric is the total number of times the target paper is cited since it was published. We use DOIs to query the Scopus API ($C_{\rm SC}$) and Crossref API ($C_{\rm CR}$), which return metadata including the citation count and the publication year. The final value is the higher citation count between them. Formally,
$$\small  C(p)=\begin{cases}
    \max\left\{C_{\rm SC}(p)\text{ and }C_{\rm CR}(p)\right\} \\
    0, & \text{otherwise}
\end{cases}$$

\paragraph{normalized\_citations\label{sec:normcite}}
This is calculated as the average number of citations per year since the target paper was published. Formally,
\begin{equation}\label{eq:nc}\small
\overline{C}(p)=C(p)/\Delta Y(p),\quad\Delta Y(p)=Y_{\rm now}(p)-Y_0(p)\end{equation}
in which $Y_{\rm now}(p)$ and $Y_0(p)$ denote the current year and the publication year of the paper. In rare cases that an API response is not available, a default value of $0$ is used. 

\paragraph{citation\_Velocity\label{sec:citation_velocity}}
Citation velocity, introduced by S2 in 2016, is an average of the publication’s citations for the last 3 years and fewer for publications published in the last year or two, which 
aims to capture the current popularity and relevance of the work \cite{kirkpatrick2016search}. This metric is pre-calculated by S2 and can be obtained by querying the S2 API using a paper identifier. 

\paragraph{citation\_next\label{sec:citation_next}}
The time window of 3--5 years after a paper is published is usually considered particularly important for measuring its impact \cite{aksnes2019citations}. This feature measures the early citation momentum of a paper. Specifically, this feature is  defined as the number of citation a paper receives in the first 3 years after its publication. Formally, 
\begin{equation}
\footnotesize \overline{C_3}(p)=\sum_{i=1}^{\Delta Y_3}c_i(p)\bigg/\Delta Y_3,\Delta Y_3=\min\left\{3,\Delta Y(p)\right\}
\end{equation}
in which $c_i(p)$ is the number of citation received in year $i$, obtained by querying the S2 API, and $\Delta Y(p)$ is defined in Eq.(\ref{eq:nc}). The year of publication is obtained from the Crossref API and Scopus API. 

\paragraph{influentialCitationCount\label{sec:influentialcount}}
Recent work has argued that not all citations are equal, e.g., \cite{Valenzuela2015IdentifyingMC}. In S2, citation metrics are calculated by an algorithm that de-emphasizes absolute citation counts, assigns differential weights to citations depending on citation context, recency, and rate to better determine level of influence. Given a paper identifier, the S2 API returns the number of \textit{influential} citations, which counts citations in which the cited paper had a strong impact on the citing work \cite{valenzuela2015identifying}. 

\paragraph{references\_count}
This metric is the number of references the target paper cites obtained from the S2 API and Crossref API, whichever is higher. We consider this feature because it reflects the extend of background and related works the current paper is based on.
We set the default value to $0$ in case of no API responses. 

\paragraph{self\_citations\label{sec:selfcite}}
Excessively citing the authors' papers can increase author's h-index, which creates a motivation to strategically use self-citation \cite{Seeber2017SelfcitationsAS} to promote the apparent impact. Self-citations has been used as a measure to complement h-index \cite{Kacem2020TrackingSI}. Intuitively, papers that self-cite disproportionately and excessively could potentially reproduce poorly.

Using the extracted author names and references for a given paper, we compute the self-citation count by excluding references authors by any co-author of the target paper. Each author name is parsed to a tuple of (last name, first name initial). Two author names match if they have the same first initial and their last names' matching score, calculated by Levenshtein distance, is above a threshold, empirically set to 85\%. The self-citation ratio is then calculated as the self-citation count divided by the total number of references. 
The accuracy of this feature depends on the quality of XML output by GROBID. Errors could be caused by author names that are not extracted from the header or bibliographic sections. By taking the GROBID extraction errors into consideration, the fuzzy matching algorithm  results in a root-mean-square-error (RMSE) of 0.09 by comparing automated and manually calculated self-citation ratios for a sample of 37 SBS papers. 

\paragraph{openaccessflag\label{sec:oa}}
Another feature considered is whether the paper has open access. 
Subscription-based access generally limits the availability of papers. The article being open access can be a potentially important features to observe. This binary feature can be obtained by querying Scopus and Crossref APIs. We assume a paper does not have open access by default. 

\paragraph{age\label{sec:age}}
This is the number of years since the paper was published

\paragraph{coCite2\label{sec:cocite}}
The co-citation index between two papers is defined as the number of papers that cite both of them. Papers with higher co-citation indices are usually highly relevant in topics. Therefore, co-citation index can be used for finding topically similar papers. For a target paper $p$, we use citation graphs to find all ``similar'' papers with non-zero co-citation indices using the S2 API. This is achieved by first finding all papers (citing papers) that cite the target paper $S_{A}=\{A_1,\cdots,A_m\}$. Then we find all references in a citing paper $A_k$: $\{r_1,\cdots,r_l\}$. We next find papers citing $r_1$: $S_B=\{B_1,\cdots,B_n\}$. The co-citation index between $p$ and $r_1$ can be calculated as $|S_A\cap S_B|$. This feature counts the numbers of similar papers within 2 years after the target paper was published.

\paragraph{coCite3\label{sec:cocite3}}
This feature is similar to coCite2 except that it counts similar papers within 3 years after the target paper was published.

\paragraph{u\_rank\label{sec:univrank}}
Intuitively, the university rank of authors can be used as a indicator of the author's accountability and credibility. We collected university ranking data from the 2020 Times Higher Education  rankings\footnote{https://www.timeshighereducation.com/world-university-rankings} and use it as a lookup table to generate the feature value. For a given paper we extract the organization the first and second author (if exists). For matching against the lookup table we use the university the first author is affiliated to. If it is not available we use the second author's affiliation. If neither author's affiliations are available, the default value is used. 
University names are normalized by removing accents, punctuation marks, and non-ASCII characters. We applied a fuzzy string matcher and set the threshold to 95\%, which achieves 100\% matching accuracy for 20 random cases with full university names. {Another lookup table mapping acronyms to full university names is used in case the latter is not available.}

Once matched, a normalized rank between 0 and 1 is calculated as $R_{\rm N}(u_i)=1-R(u_i)/100$, in which $R(u_i)$ is the ranking of university $u_i$. 
We consider only the top one hundred universities. If the university's rank is higher than one hundred, we assign $R_{\rm N}(u_i)=2$. In cases where there is no match, a default value of 2 is assigned.

\subsection{Author Features}
Features in this category are related to the authors of the target paper, obtained from S2. Author features include
\begin{itemize}[leftmargin=*]
    \item {\em author\_count}. The total number of authors of the target paper. 
    \item {\em avg\_pub}. The average number of publications of all authors of the target paper.
    \item  {\em avg\_hidx}. The average h-index of all authors of the target paper.
    \item {\em avg\_high\_inf\_cites}. The average number of highly influential citations \cite{valenzuela2015identifying} of all authors.  
    \item {\em avg\_auth\_cites}. The average number of citations of all authors. 
\end{itemize}

\begin{figure}
    \centering
    \includegraphics[width=.47\textwidth]{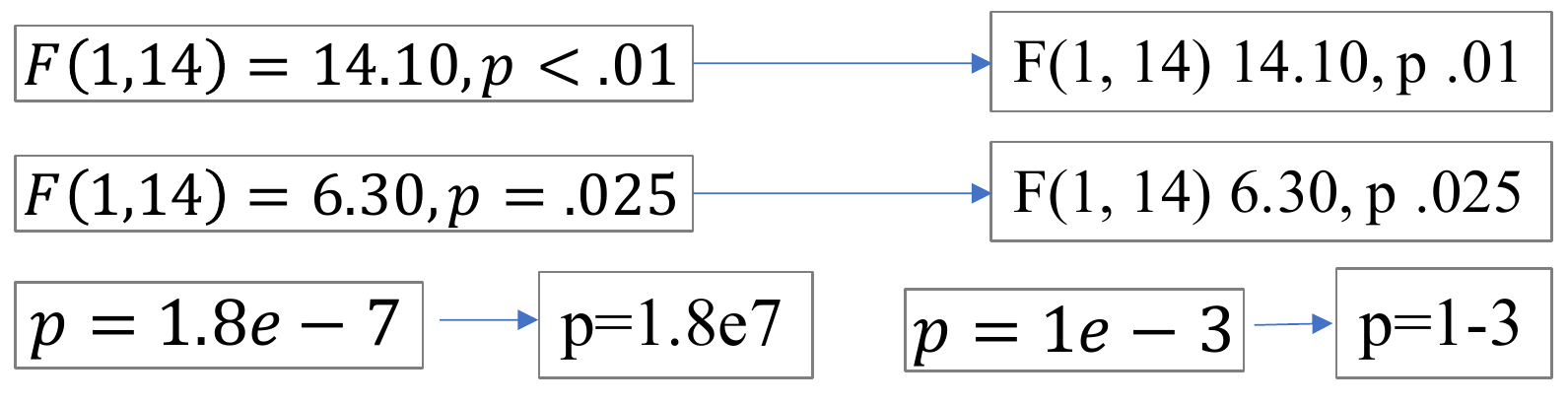}
    \caption{Typical cases in which comparison operators are missed when a PDF paper is converted to text.}
    \label{fig:pdftextcomp}
\end{figure}

\subsection{Venue Features}
Features in this category are pertaining to the conference or journal for a particular paper. All data are obtained from the Scopus API\footnote{\url{https://service.elsevier.com/app/answers/detail/a_id/14834/supporthub/scopus/}} using journal's ISSN as the identifier.

\paragraph{Venue\_CiteScore}
CiteScore was first introduced in 2016, as part of an evolving array of research metrics. The metric is a standard to help measure citation impact for journals, book series, conference proceedings and trade journals \cite{silva2020citescore}. 

\paragraph{Venue\_SNIP}
The SNIP indicator measures the average citation impact of the publications of a journal, using Scopus data. Unlike the well-known journal impact factor, SNIP corrects for differences in citation practices between scientific fields, thereby allowing for more accurate between-field comparisons of citation impact\footnote{https://www.journalindicators.com/}. SNIP is derived by taking a journal’s citation count per paper and dividing it by the citation potential in its subject field \cite{leydesdorff2010snip}. 

\paragraph{Venue\_Scholarly\_Output}
Scholarly output defines the total count of research outputs, to represent productivity. This feature is calculated as the sum of documents published in a certain venue in the 3 years prior to the current year.

\paragraph{Venue\_Percent\_Cited}
This is calculated as the proportion of documents that have received at least 1 citation.

\paragraph{Venue\_Citation\_Count}
This feature is calculated as the number of citations received in one year for the documents published in the previous 3 years.

\paragraph{SJR\label{sec:sjr}}
The SJR stands for the SCImago Journal Rank\footnote{https://www.scimagojr.com/SCImagoJournalRank.pdf}, which accounts both the number of citations received by a journal and the importance/prestige of  journals where the citations come from. It is calculated as the average number of weighted citations received in a year divided by the number of documents published in the last three years. In case of no API response, a default value of $0$ is used. 

\subsection{Statistical Features}
Statistical information is frequently reported in SBS papers when experiments are run. 
We focus specifically on p-values, a measure of the significance of the observed result. A p-value may serve as an indicator of whether findings from an experiment can be reproduced. In addition, p-values, when presented with the test statistics (e.g., t-test), are especially important references to accept or reject the null hypothesis.

\subsubsection{Extracting p-values} \label{pvaluesubsec}
To extract p-values, a PDF document is first converted to a text document. 
We compared several software packages to convert PDF to text, such as XpdfReader, PyPDF2, PDFBox, and PDFMiner and adopted {\sc pdftotext}, which produces fewer errors. This is consistent with a recent work on text extractor comparison \cite{bast2017benchmark}.

Typical errors when extracting p-value expressions include missing comparison symbols such as``$=$'', ``$>$'', and ``$<$'' (Figure~\ref{fig:pdftextcomp}), which makes the expression no longer valid. We evaluated the text converter on a random set of 37 papers (hereafter SBS37). {\sc pdftotext} successfully converted 90.1\% p-values expressions without test statistics (156 out of 173) and 82.5\% p-value expressions with test statistics (378 out of 458).

In an SBS paper, p-values can be represented with or without test statistics. The p-values without test statistics are usually in forms of
``{\tts\ p <operator> <sign><number>}'',
in which {\tts\ <operator>} is one of ``$=$'', ``$>$'', or ``$<$''. The {\tts\ <sign>} could be ``$+$'' or ``$-$'', and the {\tts\ <number>} could be an integer (e.g., {\tts\ -2}), float (e.g., {\tts\ 0.05}), or and exponential (e.g., {\tts\ 1.2e-4}). These forms can be captured by regular expressions\footnote{Regular expressions are available in the code repository.}. 

The p-values may be reported with test statistics, such as
{\tts\ t(12)=4.3, p=0.01} and {\tts\ f(21,30)=2,3, p<0.01},
which represent the result of a student's t-test and F-test, respectively. A complete list of p-values patterns in different statistical testings are tabulated in an online document\footnote{\url{shorturl.at/ghtD2}}. Using the SBS37 dataset, we compared automatically extracted p-values against the PDF and converted text. The results (Table~\ref{tab:pvalpdfeval}) indicate that our regular expressions  can capture 92\% p-values without test statistics from the original PDF, with an overall $F_1=0.792$. The precision on capturing p-values with test statistics is 0.994, with an overall $F_1=0.864$.

\begin{table}
\caption{\small Evaluation of p-value and sample size extractors against manually extracted ground truth from PDF and converted text.}\label{tab:pvalpdfeval}
\begin{center}\small
\begin{tabular}{c|c|ccc} 
\toprule
{\bf DocType} & {\bf Data Extracted} & $P$ & $R$ & $F_1$\\
\midrule
\multirow{3}{*}{PDF} & $p$-val w/ test stat &  0.695 & 0.920 & 0.792\\
                     & $p$-val w/o test stat& 0.994 & 0.765 & 0.864\\
                     \cmidrule{2-5}
                     & Sample size & 0.592 & 0.990 & 0.741\\
                     \midrule
\multirow{3}{*}{TXT} & w/ test stat & 0.698 & 0.985 & 0.817\\
                     & w/o test stat & 0.994 & 0.926 & 0.959\\
                     \cmidrule{2-5}
                     & Sample size & 0.592 & 1.000 & 0.743 \\
\bottomrule
\end{tabular}
\end{center}
\end{table}

\subsubsection{Derived Features From p-values}
\begin{itemize}[leftmargin=*]
    \item {\em real\_p}. A p-value less than $0.05$ is usually regarded as a relatively high confidence to exclude the null hypothesis. Because we do not distinguish each hypothesis test, the minimum p-value among all the p-values extracted is used as this feature. 
    \item {\em real\_p\_sign}. The signs parsed from p-value expressions. The ``<'', ``='', and ``>'' are encoded as $-1$, $0$, and $1$, respectively.
    \item {\em p\_val\_range}. The p-value range is obtained as the difference of the highest and the lowest p-value in the paper.
    \item {\em num\_hypo\_tested}. We assume the number of hypothesis tests is equal to the total number of p-values with test statistics. 
    \item {\em extend\_p}. A Boolean indicating whether the p-value features are associated with a test.
    \item {\em num\_significant}. This metric is calculated as the total number of significant p-values ($\leq0.05$) including those with and without test statistics as recognized by our parser. 
    \item {\em sample\_size} \label{samplesizeeqs} In an SBS experiment, the sample size is defined as the number of participants or observations. The sample size may explicitly appear in the paper text or can be derived from the p-value test statistic expressions. In one scenario, the sample size could be represented as a integer in free text and is usually noted as  $N=N_0$ or $n=n_0$, in which $N_0$ and $n_0$ are integers. In another scenario, the sample size can be parsed out by matching the $N=N_0$ pattern in a p-value expression (such as seen in the Chi-squared test). If the $N=N_0$ pattern is missing, the second argument inside $\chi^2$ is treated as the sample size. For certain tests, the sample size can be computed from the test statistic expressions. For example, if a t-test expression is \\
    {\tts\ t(df)=number, p<number}, \\
    then the sample size is {\tts\ df+1}. 
    
    The sample size extractor is evaluated using the SBS37 corpus, which achieves a high recall (0.990 for PDF and 1.000 for text) but relatively low precision (0.592). The extractor can be improved using the context around an expression to decide whether it includes a sample size. 
\end{itemize}

\subsection{Semantic Features}
\paragraph{Citation and Reference Intents}
A paper could be cited for different reasons. To account for the citation intent, S2 calculates the number of times a given paper is cited as background, methodology, or result \cite{jurgens2018measuring}. Similarly, citation intent can be obtained for references cited in the given paper. This generates 6 features, namely, {\em reference\_background}, {\em reference\_methodology}, {\em reference\_result}, {\em citations\_background}, {\em citations\_methodology}, and {\em citations\_result}. 

\paragraph{upstream\_influential\_methodology\_count} 
This feature is the number of papers referenced in the target paper in which the citation context is classified as methodology and the referenced paper was classified as influential by S2.

\paragraph{funded\label{sec:ackextract}}
Acknowledgements are ubiquitous in research papers. 
We consider acknowledgement of a funding agency is a factor for predicting the reproduciblity. We extract acknowledgement organizations using {\sc AckExtract}, a framework that distinguishes mentioned and acknowledged entities in a paper \cite{wu2020ackextract}. {\sc AckExtract} classifies sentences, recognizes all people and organizations  from acknowledgement sentences, and then differentiate between acknowledged and mentioned entities. {\sc ActExtract} was evaluated using a corpus of 100 acknowledgement paragraphs containing 146 PEOPLE and 209 ORGANIZATION entities and achieved an overall $F_1$=0.92. 

\paragraph{subject and subject\_code\label{sec:subject}}
In Elsevier, serial titles are classified into 335 \emph{subject fields} by human experts under the All Science Journal Classification (ASJC) scheme. Each subject field is associated with a code ranging from 1000-3700, belonging to 5 \emph{subject areas} -- Multidisciplinary, Life Sciences, Social Sciences \& Humanities, Physical Sciences, and Health Sciences. We encode the subject field and the subject area returned by the Elsevier Serial Title API into features named \emph{subject} and \emph{subject\_code}. 

\section{Reproducibility Prediction}
One important question we attempt to answer is whether it is possible to predict reproducibility from features that can be directly extracted from the paper without redoing the experiments. One existing work used prediction markets to forecast the results of novel experimental designs that test established theories \cite{viganola2021using}. Another recent study used supervised machine learning models and captured the differences in $n$-grams between replicating and non-replicating paper. In this study, we treat it as a classification problem and use 10 supervised machine learning models to classify each paper into one of two categories: reproducible vs. non-reproducible \cite{yang2020pnas}. We tried 10 different classifiers to reduce the potential disadvantage of certain classifiers due to the relatively small sample size. Furthermore, we investigate the possibility  to reduce the dimensionality of the feature space by removing strongly correlated features. 

\begin{table}
    \centering
      \caption{Subject distribution of our dataset.}
    \begin{tabular}{c|l}
    \toprule
    {\bf Count} & \multicolumn{1}{c}{\bf Subject}\\
    \midrule
    64  & Psychology \\
    30     & Sociology and Political Science \\
  22 & Linguistics and Language\\
  5 &  Social Psychology\\
  5 & Psychological Science \\
 2 & Clinical Psychology\\
 2 & Arts and Humanities\\
 2 & History and Philosophy of Science\\
 1 & Psychiatry and Mental Health\\
 1 & Behavioral decision making \\ 
 1 & Philosophy\\
 1 & Social Sciences \\
 1 & Strategy and Management \\
 1 & Developmental Neuroscience \\
 1 & Cognitive Neuroscience \\
 \midrule
 {\bf 139} & {\bf Total}\\
 \bottomrule
  \end{tabular}
  \label{tab:papers}
\end{table}
\subsection{Data}
We construct a corpus of {139} SBS papers collected from three sources, covering a broad spectrum of subjects (Table~\ref{tab:papers}).  
These papers have been under careful examination and their reproducibility has been manually labeled by domain experts. 

The reproducibility project \cite{osc2015estimating} replicated 99 experimental studies published in three reputable psychology journals (Psychological Science, Journal of Personality and Social Psychology, Journal of Experimental Psychology: Learning, Memory, and Cognition). The results from these replications have been labeled as either replicated or non-replicated and added to our dataset. We also included 12 replication studies from Many Labs 1 \cite{Klein2014} and 28 replication studies from Many Labs 2 \cite{Klein_Vianello_2019} project. This resulted in a total of 139 labeled papers. A portion of papers used in our work were adopted in a recent reproducibility study \cite{yang2020pnas}. 

\begin{figure}
    \centering
    \includegraphics[width=0.74\textwidth]{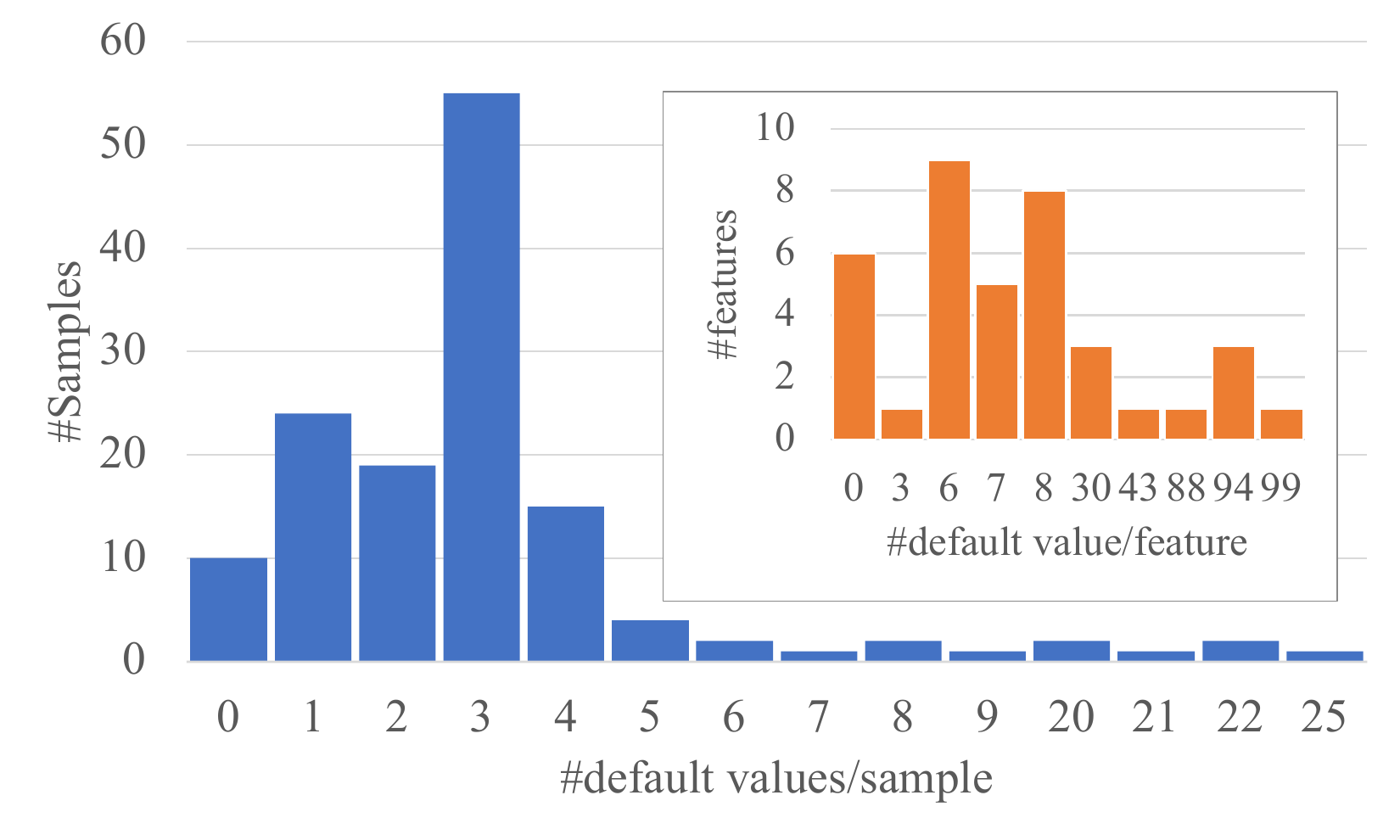}
    \vspace{-3mm}
    \caption{Distribution of the number of default values for each sample (blue), and the number of default values for each feature (orange).}
    \label{fig:defaultfrac}
\end{figure}

\subsection{Feature Extraction}
We applied the FEXRep framework to papers in our ground truth dataset and extracted 41 features. As mentioned in Section~\ref{sec:framework}, we apply default values for certain features if the real values are unavailable. To mitigate the artifact introduced by default feature values, we exclude features if there are less than 15 samples with real values, including real\_p\_sign (5 samples), p\_val\_range (14 samples), and sample\_size (14 samples). Figure~\ref{fig:defaultfrac} displays the number of samples as a function of default values and the distribution of default values over features. This figure indicates that most samples (96\%) have less than 10 features with default values and most features (84\%) have less than 30 samples with default values. We refer these 38 features as {\bf core features}. 

\subsection{Classification Using Core Features}
\begin{figure}
    \centering
    \includegraphics[width=0.74\textwidth]{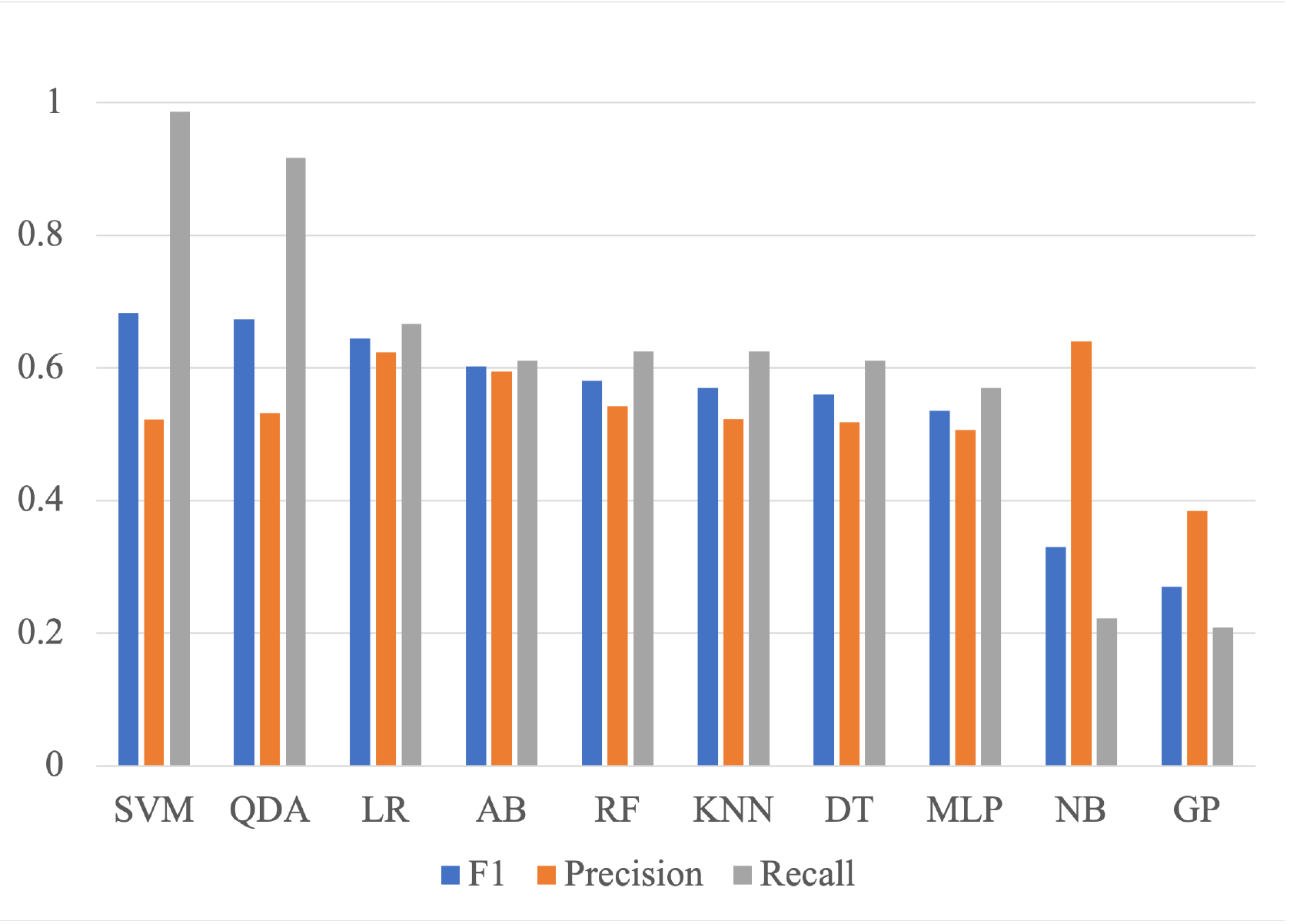}
    \caption{Five-fold CV results using core features sorted by F1 from left to right. }
    \label{fig:mlresults5fcv}
\end{figure}

In this section, we apply supervised machine learning models to predict whether a paper is reproducible or not using core features. Our goal is to find out how the prediction accuracy depends on classic machine learning models. The following models were applied. 
\begin{enumerate}[leftmargin=*]
    \item Logistic regression. 
    \item K-nearest neighbors ($k$-NN), in which $k$ is set 5.  
    \item Gaussian process classifier, a non-parametric supervised probabilistic classifier, which assumes that all random variables follow Gaussian distributions. We applied the radial basis function (RBF) as the kernel. 
    \item Decision tree. Gini impurity is the splitting criterion.
    \item Random forest. The max depth is set to 2. The number of estimators is set to 200, and Gini impurity is the splitting criterion. 
    \item Multilayer perceptron (MLP). MLP is a neural network based supervised classifier that can learn non-linear models. 
    \item AdaBoost (AB). AB is an ensemble model that fits a sequence of weak learners (i.e., models that are only slightly better than random guessing) on repeatedly modified versions of the data. 
    \item Na\"{i}ve Bayes (NB). NB is a probability classifier based on Bayes’ theorem with the assumption of conditional independence between every pair of features given the value of the class variable. 
    \item Quadratic Discriminant Analysis (QDA). QDA uses quadratic surfaces to divide sample points in the feature space. 
    \item Support vector machine (SVM). The RBF kernel was applied. 
\end{enumerate}
Because the sample size is relatively small, we apply a five-fold cross validation (CV) for all models. Figure~\ref{fig:mlresults5fcv} shows that evaluation results using the  core features exhibit significantly different performances. The highest $F_1$=0.68 is achieved by SVM and QDA, followed by LR with $F_1$=0.64 and AB with $F_1$=0.60. SVM and QDA also achieve superior recalls with $R$=0.99 and $R$=0.92, respectively. The highest precision is achieved by NB with $P$=0.64. 

\section{Feature Analysis}
\subsection{Correlations Between Features}
\label{sec:correlation}
\begin{figure}
    \centering
    \includegraphics[width=0.74\textwidth]{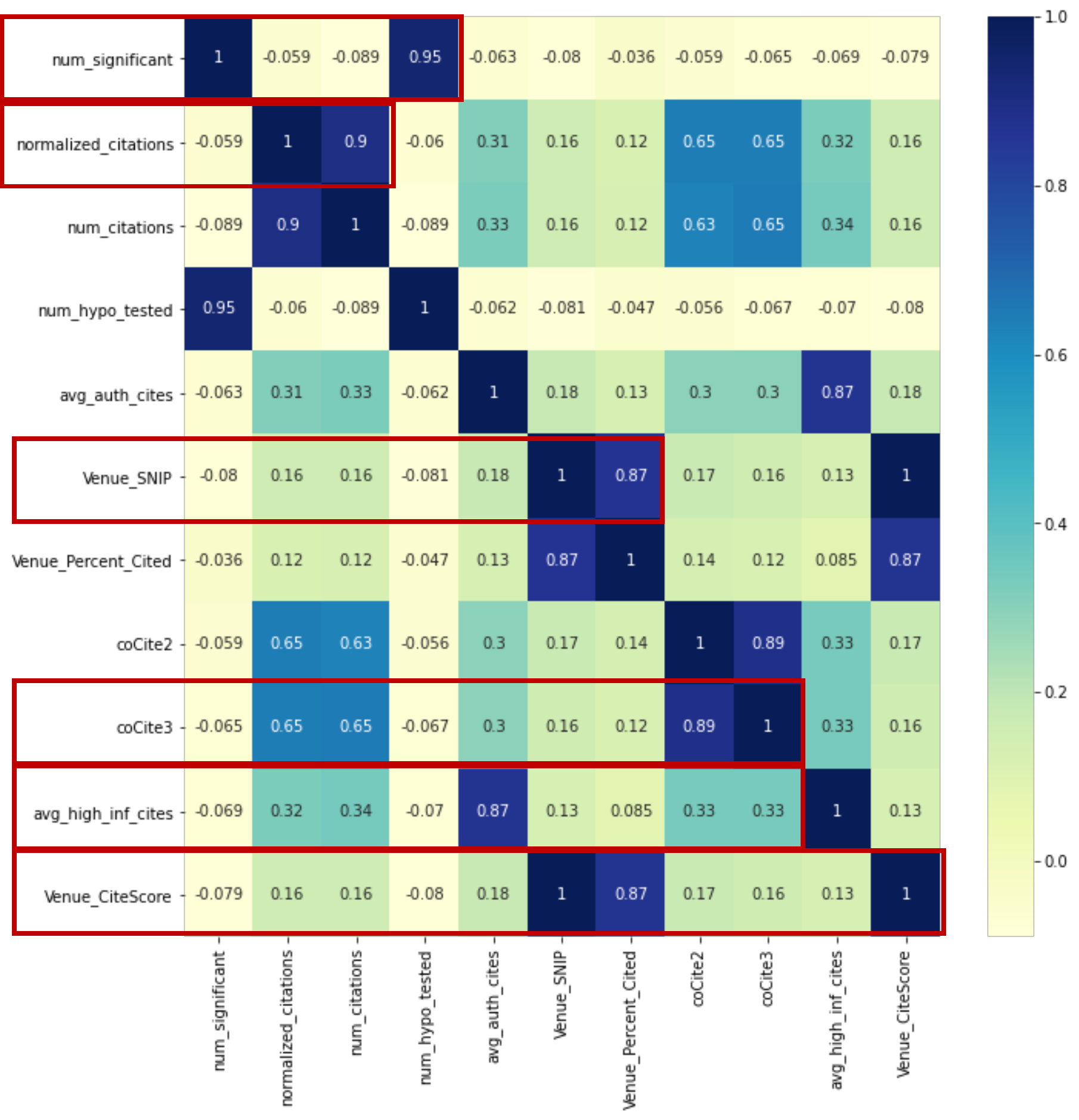}
    \caption{Kendall's $\tau$ matrix between highly correlated features. Excluded features are enclosed by red boxes. Note that the matrix is symmetric, so we only study numbers in the lower left triangle.}
    \label{fig:kendalltau}
\end{figure}

The FEXRep framework was designed to extract features that are potentially useful for reproducibility prediction. However, certain features could be correlated with each other, making them less useful in prediction. To capture these correlations, we calculate the Kendall's $\tau$ coefficient \cite{kendall1938new} between any two features with continuous values in our list (features with categorical values are excluded for this analysis). Kendall's $\tau$ coefficient ranges from $-1$ to $+1$. A stronger correlation between two variables results in a higher absolute value. Two random variables without any correlation has $\tau=0$. We determine feature $i$ and $j$ to be strongly correlated if $\tau_{i,j}>0.8$. We drop the feature with less real data available. We excluded 6 features that are strongly correlated with at least another feature (Figure~\ref{fig:kendalltau}), including 
\begin{itemize}[leftmargin=*]
    \item {\em num\_significant} (correlated with num\_hypo\_tested). \item {\em normalized\_citations} (correlated with num\_citations)
    \item {\em Venue\_SNIP} (correlated with Venue\_percent\_cited)
    \item {\em coCite3} (correlated with coCite2) 
    \item {\em avg\_high\_inf\_cites} (correlated with avg\_auth\_cite)
    \item {\em Venue\_CiteScore} (correlated with Venue\_percent\_cited)
\end{itemize} This results in 33 features that are relatively independent with each other. We refer them as {\bf reduced features}. Using \emph{reduced features}, we classified the samples using all machine learning models and obtained consistent results in general (Figure~\ref{fig:norm}). 

\subsection{Feature Selection}
We apply two methods to identify the most relevant features to the reproducibility label: ANOVA-F and Mutual Information (MI). 

\paragraph{ANOVA-F} ANOVA (analysis of variance) is a parametric statistical hypothesis test for determining whether the means from two or more samples of data come from the same distribution or not. F-test is a class of statistical tests that calculate the variance from two different samples. 
ANOVA uses F-test to determine whether the variability between group means is larger than the variability of the observations within the groups \cite{kuhn2019feature}. The larger the value, the more useful the feature is in classification. 

\paragraph{Mutual Information (MI)} MI measures the amount of information one can obtain from one random variable given another. The MI between two variables $x$ and $y$ can be calculated as $I(x,y)=H(x)-H(x|y)$, in which $H(x)$ is the entropy for $x$ and $H(x|y)$ is the conditional entropy for $x$ given $y$. The smaller the value is, the more independent the two variables are. We use non-parametric methods based on entropy estimation from k-nearest neighbors distances to calculate MI \cite{kraskov2004estimating,ross2014mutual} . MI values should be non-negative. 

\subsection{Selecting Top Features\label{sec:topfeatures}}
\begin{figure}
    \centering
    \includegraphics[width=0.74\textwidth]{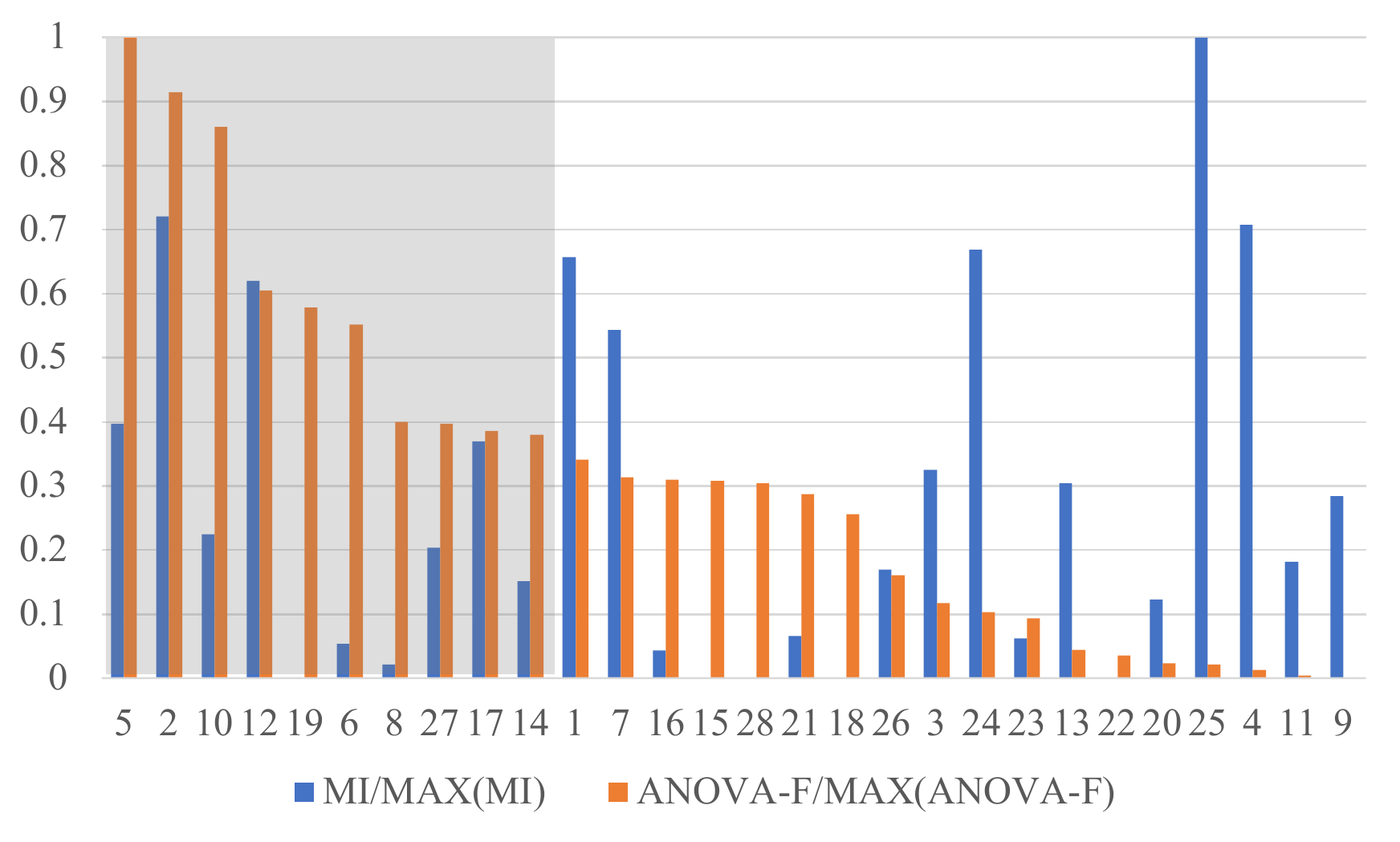}
    \caption{Distribution of mutual information (normalized) and ANOVA-F (normalized) values of \emph{reduced features}. The x-axis labels are core feature IDs used in these calculations. }
    \label{fig:mianovaf}
\end{figure}

We compared  scores calculated using ANOVA-F and MI and show the distributions in Figure~\ref{fig:mianovaf}. We normalized both scores by dividing each value by the maximum value. The shaded region shows top features indicated by ANOVA-F scores (the larger the better). The top features and normalized MI and ANOVA-F scores are tabulated in Table~\ref{tab:topfeatures}. 

\setlength{\tabcolsep}{3pt}
\begin{table}
    \centering
    \caption{Top 10 features identified by MI (top portion) and ANOVA-F (bottom portion). Feature IDs correspond to x-labels in Figure~\ref{fig:mianovaf}. We show their normalized MI and ANOVA-F values.  Blue text are cross-listed features by both MI and ANOVA-F.}
    \begin{tabular}{cclccc}
    \toprule
    {\bf ID} & &{\bf Feature}     &  {\bf MI} & {\bf ANOVA-F} \\
    \midrule
    22   &  & Venue\_citation\_count & 0  & 0.036 \\
    18   &  & coCite2 & 0 & 0.255  \\
    28   &  & age & 0 & 0.305 \\
    15 & &citations\_result & 0 & 0.308  \\
    19 & &{\color{blue}num\_hypo\_tested} & 0 & 0.578\\
    8 & &{\color{blue}influentialCitationCount} & 0.021 & 0.340 \\
    16 & \greendot{2pt} & citations\_methodology & 0.043 & 0.310  \\
    6 & & {\color{blue}upstream\_influential\_methodology\_count} & 0.053 & 0.552 \\
    23 & & Venue\_Scholarly\_Output & 0.062 & 0.093  \\
    21 & & extend\_p & 0.066  & 0.287 \\
    \midrule
    5 & \greendot{2pt} & self\_citations & 0.397 & 1.00\\
    2 & \greendot{2pt} & author\_count & 0.721 & 0.915 & \\
    10 & \greendot{2pt} & influentialReferencesCount & 0.224 & 0.860\\
    12 & \greendot{2pt} & reference\_result & 0.620 & 0.605 \\
    19 & \greendot{2pt} & {\color{blue}num\_hypo\_tested} & 0 & 0.578\\
    6 & \greendot{2pt} & {\color{blue}upstream\_influential\_methodology\_count} & 0.053 & 0.552\\
    8 & \greendot{2pt} & {\color{blue}influentialCitationCount} & 0.021 & 0.340 \\
    27 & \greendot{2pt} & avg\_auth\_cites & 0.204 & 0.397 \\
    17 & & citations\_next & 0.370 & 0.385 \\
    14 & & citations\_background & 0.152 & 0.380 \\
    \bottomrule
    \end{tabular}
    \label{tab:topfeatures}
\end{table}

The MI and ANOVA-F select different sets of relevant features. This is because these two methods captures different types of relations. The ANOVA-F captures linear relationships between variables and the MI captures any types of relationships.  As seen in Table~\ref{tab:topfeatures}, there are 3 cross-listed top 10 features identified by both ANOVA-F and MI (in blue text). Among them, the num\_hypo\_tested feature has MI=0. Estimates of MI can result in negative values due to sampling errors, and potential violation in the assumption that sample rate is high enough for point density to be locally uniform around each point. In our implementation\footnote{\url{https://scikit-learn.org/stable/modules/generated/sklearn.feature_selection.mutual_info_classif.html}}, negative MI values are cast to zero. Because of this, we do not select them as top features and focus on features ranked by ANOVA-F. 

\begin{figure}
    \centering
    \includegraphics[width=0.74\textwidth]{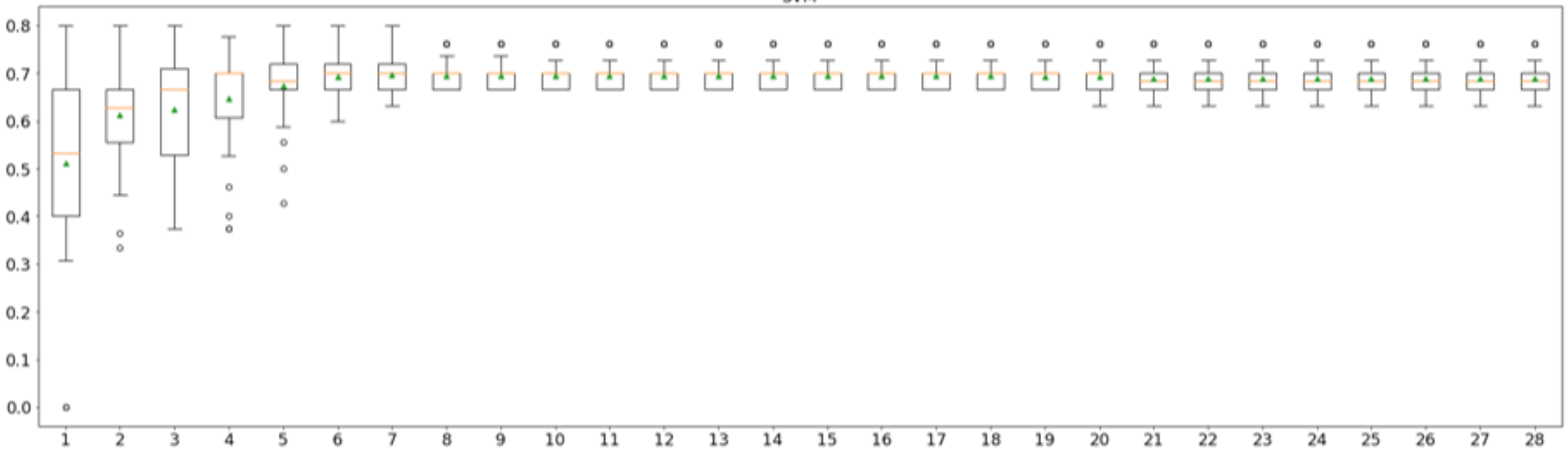}
    \caption{Box-and-whisker plot of SVM F1-measures for each number of selected features identified using ANOVA-F. The green triangles represent arithmetic means. The open dots are outliers beyond caps. The red short lines show medians. The x-labels are feature serial numbers and not feature IDs.}
    \label{fig:whiskerboxanovaf}
\end{figure}

To investigate how the model performance changes with top features, we evaluate model configurations on classification tasks using repeated stratified 5-fold CV. We choose SVM as the classifier and incrementally add more relevant features selected by ANOVA-F. The box-and-whisker plots are shown in Figure~\ref{fig:whiskerboxanovaf}.  The classifier achieves almost the best performance with the top 8 features identified using ANOVA-F. Adding more features do not seem to help. In fact, adding the last 7 features marginally decreases the performance. We then identify the top 8 features as the most relevant features. To be conservative, we add \emph{citations\_methodology} identified using MI as another relevant feature. These 9 features are marked with a orange dot in Table~\ref{tab:topfeatures}. 

\begin{figure}
    \centering
    \includegraphics[width=0.74\textwidth]{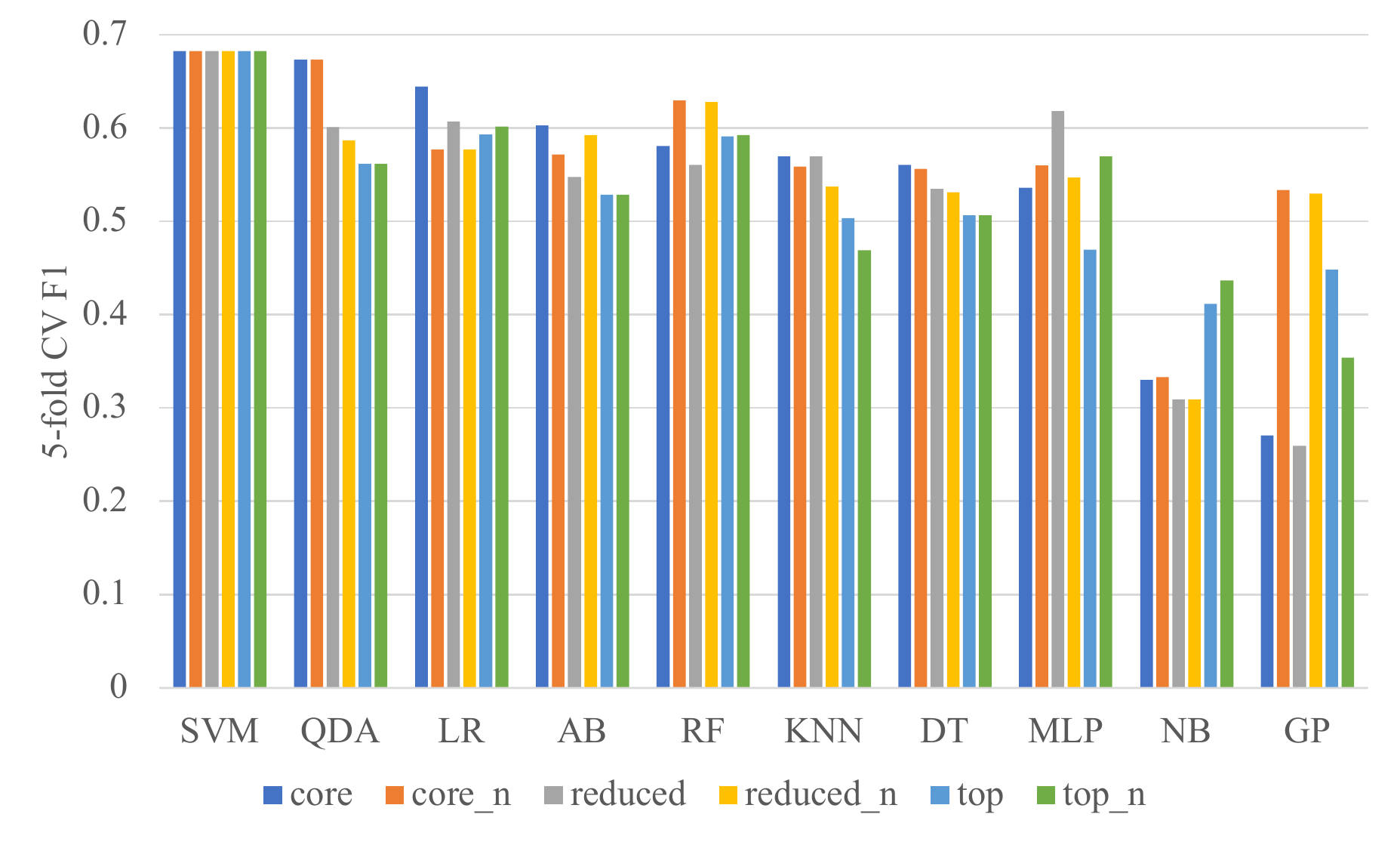}
    \caption{Comparison of F1-measures of classifiers trained on core features, normalized core features, reduced features, normalized reduced features, top 9 features, and normalized top 9 features.}
    \label{fig:norm}
\end{figure}
\subsection{Classification with Top Features}
To verify top features identified above produce consistent performance across other classifiers, we run the 5-fold CV on all classifiers using the top 9 features selected and compare the performances with classification results using core features and reduced features. We also compare the performance with and without feature normalization. To normalize a feature, we scale a feature to a range between 0 and 1. The transformation is given by $X^\prime=(X-X_{\rm min})/(X_{\rm max}-X_{\rm min})$. The comparison results are illustrated in Figure~\ref{fig:norm}. The F1 of SVM is stable with a marginal decrease with reduced and top features. QDA, LR, AB, KNN, and DT exhibit a general decline when trained with reduced features and top-9 features. Normalizing features helps boosting performances in certain conditions. In RF, MLP, NB, and GP, classifiers trained on the top-9 features outperform the core and reduced features. Normalizing feature may or may not boost the performances. 
Except for QDA, which exhibits a drop of $\sim0.11$ and KNN, which exhibits a drop of $\sim0.07$, the other classifiers either show a mild decrease ($<0.05$) or an increase between core features and top-9 features. The increase of F1 with top-9 features indicate that the classifier may overfit when trained with core features or reduced features, which can be mitigated by adding more training samples. Overall, Figure~\ref{fig:norm} verifies that the top-9 features selected in Section~\ref{sec:topfeatures} produce generally consistent results  across most classifiers except for QDA and KNN. Feature normalization helps to boost performances in certain cases.  


\section{Conclusion and Discussion}
In this work, we develop a framework called FEXRep, which automatically extracts 41 features from SBS research papers. The framework was designed to be modular, scalable, and customizable. New extractors can be added and existing extractors can be updated for better performance. We then use extraction results of this framework to predict the reproducibility of a corpus of {139} SBS papers. By conducting statistical correlation and feature analysis, we finally selected 9 top  features, which we believe to be most important. Our work sheds light on the power of using classic machine learning models for evaluating research claims. The normalized top-9 features achieved the best $F_1$=0.68 using SVM, the best precision of 0.69 using QDA, and the best recall of 0.99 using SVM. 

Our study has two limitations. The first is the relatively small sample size. Unfortunately, determining the reproducibility of a claim within a research paper usually requires tremendous effort, rich domain knowledge, and close collaboration, e.g., \cite{camerer2016science}. With the advocacy and adoption of open science, more papers will be labeled by domain experts, e.g., the repliCATS project \cite{hanea2021mathematically,nosek2021replicability}, and the prediction model will be more robust.

Another limitation is caused by missing values which were set to default values. Lots of default values make us underestimate the true variance of a feature. Most predictive modeling algorithms cannot handle missing data. One simple mitigation is  imputing the median for continuous and the modus for discrete predictors. More sophisticated methods to handle missing data build prediction models that estimate missing data.

\bibliographystyle{unsrt}  
\bibliography{references}  

\end{document}